\documentclass[conference]{IEEEtran}
\IEEEoverridecommandlockouts
\usepackage{cite}
\usepackage{bm}
\usepackage{amsmath,amssymb,amsfonts}
\usepackage{algorithmic}
\usepackage{graphicx}
\usepackage{caption}
\usepackage{textcomp}
\usepackage{xcolor}
\usepackage{stfloats}
\def\BibTeX{{\rm B\kern-.05em{\sc i\kern-.025em b}\kern-.08em
    T\kern-.1667em\lower.7ex\hbox{E}\kern-.125emX}}
\begin{document}

\title{WindMill: A Parameterized and Pluggable CGRA Implemented by DIAG Design Flow
\thanks{Identify applicable funding agency here. If none, delete this.}
}

\author{
	Haojia Hui$^{1}$\textsuperscript{$\dag$}, Jiangyuan Gu$^{1 2}$\textsuperscript{$\dag$}$^{\ast}$, Xunbo Hu$^{1}$, Yang Hu$^{1}$, Leibo Liu$^{1}$, Shaojun Wei$^{1}$ and Shouyi Yin$^{1 2}$$^{\ast}$ \\
	1. School of Integrated Circuits, Tsinghua University, Beijing, China, 100084 \\
	2. International lnnovation Center of Tsinghua University Shanghai Integrated Circuit Research Platform,\\ Shanghai, Chine, 200062}


\maketitle

\begin{abstract}
	With the cross-fertilization of applications and the ever-increasing scale of models, the efficiency and productivity of hardware computing architectures have become inadequate. This inadequacy further exacerbates issues in design flexibility, design complexity, development cycle, and development costs (4-d problems) in divergent scenarios. To address these challenges, this paper proposed a flexible design flow called DIAG based on plugin techniques. The proposed flow guides hardware development through four layers: definition(D), implementation(I), application(A), and generation(G). Furthermore, a versatile CGRA generator called WindMill is implemented, allowing for agile generation of customized hardware accelerators based on specific application demands. Applications and algorithm tasks from three aspects is experimented. In the case of reinforcement learning algorithm, a significant performance improvement of $2.3\times$ compared to GPU is achieved.
\end{abstract}


\section{Introduction}\label{sec:introduction}
With the continuous improvement of algorithms, researchers in various industries and academia are actively exploring their utilization in solving large-scale problems and discovering new techniques. However, the integration of complex algorithms and the expansion of application domains pose significant challenges to hardware architecture design in two dimensions. In the spatial dimension, the cross-fertilization of different application domains introduces complexity to algorithm implementation and increases the scale of computing. Simultaneously, in the temporal dimension, the hardware platform must keep pace with the rapid development of algorithms. These challenges result in fractional and complex application scenarios. \textbf{ Consequently, they further exacerbate the issues of design flexibility, design complexity, development cycle, and development cost, collectively named as the \textit{4D-Problem}, for hardware accelerator architectures.}

\begin{figure}[htbp]
	\begin{center}
		\includegraphics[width=3.4in]{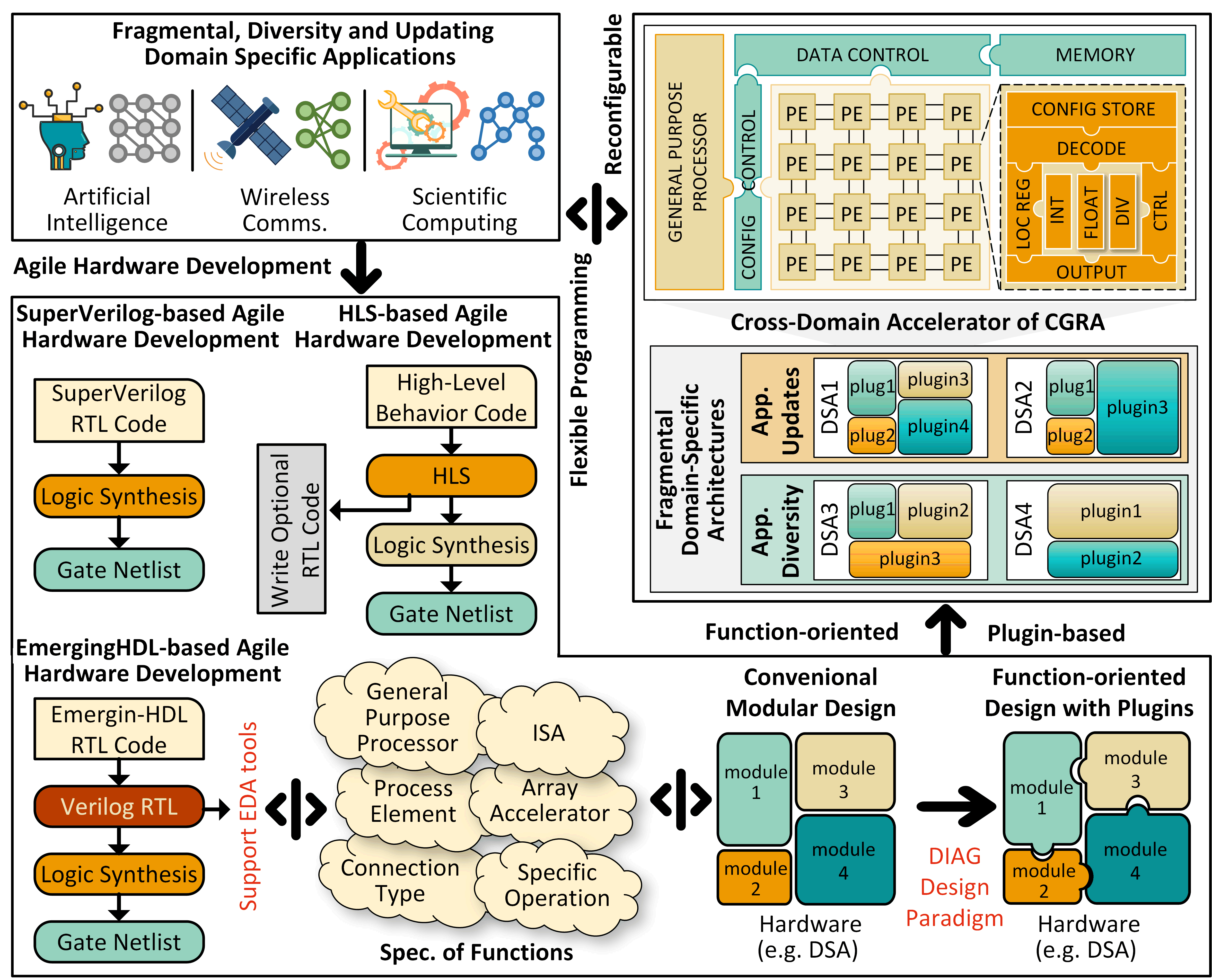}
	\end{center}
	\vspace{-0.2cm}
	\caption{Agile hardware design flow for fragmental, diverse and updating domain-specific applications}
	\vspace{-0.7cm}
	\label{fig:intro}
\end{figure}

Traditional domain-specific architectures (DSAs) show inadequate flexibility in versatile application scenario, resulting in limitation in PPA for different algorithms. To address challenges in spatial dimension, configurable spatial architecture like coarse grained reconfigurable architecture (CGRA) gains more attention with efficient computing resources utilization and powerful cross-domain application mapping ability, as shown in Fig.\ref{fig:intro}.\textbf{However, because of long development cycle and expensive cost caused by architecture diversity in computing units and connection types, the mitigation of \textit{4D-Problem} that CGRA brings is limited.}

In addition, to address the challenges in temporal dimension, the agile design methodology has been introduced into the hardware development process. Fig.\ref{fig:intro} illustrates three mainstream approaches to agile hardware development. \textit{i) SuperVerilog-based hardware design}, which improves the development platform by introducing more abstract syntax into conventional Hardware Description Languages (HDLs) such as SystemVerilog. \textit{ii) HLS-based hardware design}, which applies High-Level Synthesis (HLS) into hardware circuits scheduling using languages like C++\cite{ref:hls}. \textit{iii) EmergentHDL-based hardware design}. These HDLs are specifically developed to improve hardware design productivity and keep up with the rapid evolution of application scenarios, spanning from modeling to emulation and from generators to system integration. \textbf{They introduce an abstraction hierarchy, provide coherence, and support sophisticated EDA tools.} As a result, this paper focuses on exploring the benefits of Emergent HDLs.

Currently, the emergent HDL-based agile development researches predominantly focus on the following three aspects:
\begin{itemize}
	\item Developing agile HDLs with abstraction syntax. 
	
	\item Constructing configurable hardware architectures.
	
	\item Improving the development toolchains. 
\end{itemize}

To adapt to changing data-flow workloads and improve the \textit{4D-Problem} in CGRA hardware development, this paper realized a parameterized and pluggable CGRA called WindMill, which adopts an agile design flow following with the steps through \textit{Definition}, \textit{Implementation}, \textit{Application}, and \textit{Generation} (DIAG). The contribution paper are as follows:

\emph{1) \textbf{A versatile CGRA generator is implemented using the DIAG design flow.}} A pluggable bottom-up strategy that utilizes the \textit{Function-Plugin-Service} approach is presented. This transforms the generation of versatile cross-domain CGRAs with diverse application scenarios.

\emph{2) \textbf{The DIAG agile design flow is concluded under SpinalHDL.}} As the guidence of productive hardware development procedure, it satisfies the demand of parametric computing architecture, fine-grained heterogenous integration and further able to generate efficient pluggable circuits.

\emph{3) \textbf{Experiments demonstrating comprehensive computing capability in cross-domain scenarios are conducted.}}  The results from divergence algorithms shows adequate data-flow flexibility and system-level applications presents higher computing performance.


\section{Background} \label{sec:background}

Under the challenges from both spatial and temporal dimensions, both industry and academic proceed to redefine the design approach to diminish the 4-d problem. Although customized chips provide favorable PPA trade-offs, the prohibitive costs of NRE are undesirable. Therefore, a template—chip generator\cite{ref:generator} that can generate specialized architecture under different goals or constraints is up-and-coming. Research efforts in the trend of chip generator invest in the following two aspects:

\emph{1) \textbf{Hardware Development Agility:}} As advancements in software programming techniques continue, efforts to improve the productivity of hardware development persist. These efforts have evolved from hardware description design concepts to the extension of abstraction syntax into hardware description languages (HDLs), such as SystemVerilog, and from modular design methodologies to structured programming. Among the three hardware design approaches discussed in Section \ref{sec:introduction} (SuperVerilog-based, HLS-based, and Emergent HDL-based), this paper focuses on the Emergent HDL-based approach due to its superior abstraction capabilities derived from embedded languages (e.g., Scala) and full support for electronic design automation (EDA) tools based on VHDL/Verilog. In the Emergent HDL-based design, mainstream platforms can be categorized into imperative-based platforms like Python (e.g., MyHDL, PyRTL \cite{ref:pymtl}), and functional-based platforms like Scala (e.g., Chisel \cite{ref:chisel}, SpinalHDL \cite{ref:spinalhdl}). The latter group has gained a wider audience due to the stability features. Remarkable state-of-art projects have been developed using this approach, including Rocket-chip, BOOM, VexRiscv \cite{ref:vex}, among others. However, these works rarely delve into the improvements to conventional design strategies and methods on the flow of designing an agile generator framework. Although VexRiscv \cite{ref:vex} achieves fine-grained function reconfiguration and out-of-order hardware elaboration using plugin techniques, the author did not extract and extend this method to a broader hardware architecture.

\emph{2) \textbf{Hardware Architecture Flexibility:}} Hardware accelerators commonly require customization due to achieve significant PPA improvements. For instance, TPUs\cite{ref:tpu} adopt systolic array and unified buffer based on computing features derived from deep learning algorithms. This gains $15~30\times$ performance improvement and $30~80\times$ energy efficiency compared to contemporary GPU architectures. However, this significant progress is reached with the cost of programmability. To address this, CGRA\cite{ref:cgra} architecture is proposed. Initially, the typical representation of CGRA was RAW, a 2D array of RISC-style pipelined functional units (FUs) that communicate via a mesh network. Subsequently, MorphoSys \cite{ref:mosys} pioneered the SoC integration of CGRA and a host processor. Nowadays, Plasticine \cite{ref:plasticine}, an architecture from Stanford, realizes a parametrizable CGRA generator using Chisel.  The evolution of CGRA  progressed towards flexible programmability, fine-grained integration, and parameterized generators. However, quantitative parameterized architecture is limited in application scenario transformation. Generators facilitating extension of blocks on basic framework, like Lego, is promising.

\subsection{Motivation}

\begin{figure}[t]
	\begin{center}
		\includegraphics[width=3.4in]{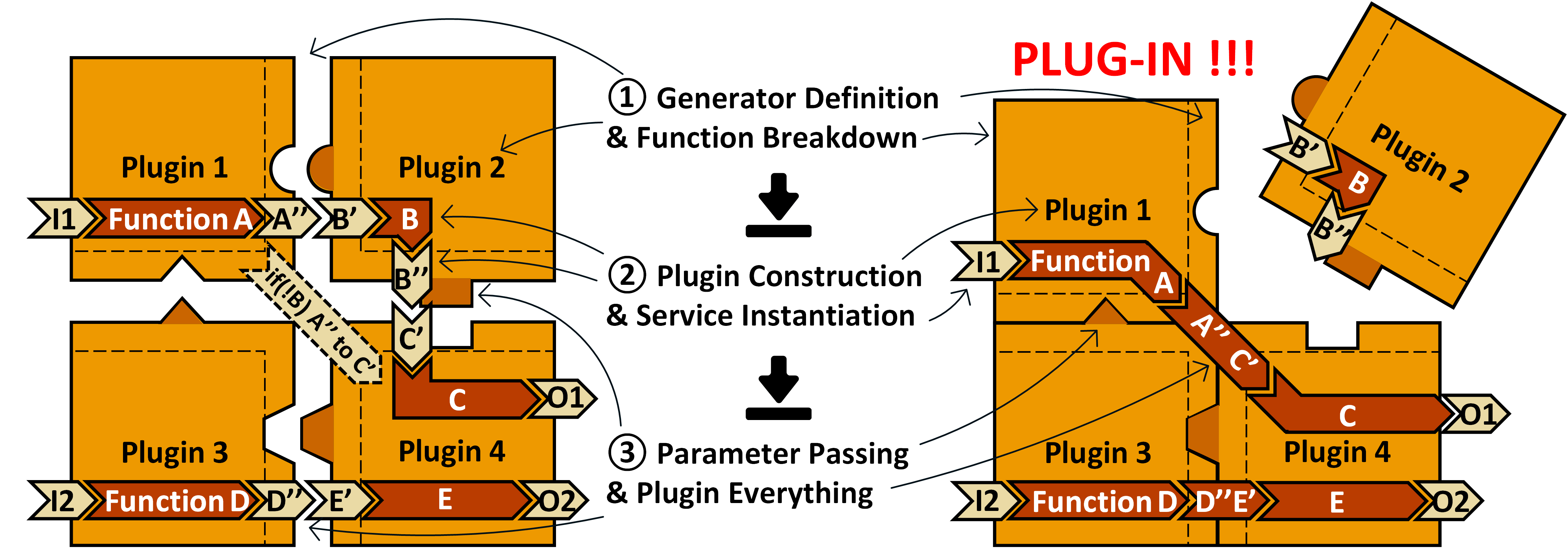}
	\end{center}
	\vspace{-0.4cm}
	\caption{Comparison between Modular and Functional}
	\vspace{-0.6cm}
	\label{fig:motivation}
\end{figure}

\textbf{Therefore, this paper implements WindMill CGRA with DIAG (Sec-\ref{sec:paradigm}) design flow aiming to bridge the gap between hardware architecture flexibility and hardware development agility.}

\section{DIAG Design Flow} \label{sec:paradigm}

From challenges described in Sec-\ref{sec:introduction} and development trends presented in Sec-\ref{sec:background}, four generic demands are concluded for agile hardware generator:

\begin{itemize}
	\item Agile Hardware Development Language.
	\item Parameterized Architecture Definition.
	\item Pluggable Chip Generator.
	\item Compatible Simulation Toolchain. 
\end{itemize}

However, hardware description differs from software programming in terms of concurrency, stationary, and physicality. A language which is characterized with multi-thread schedule and const arguments optimization is essential for constructing hardware generators, like SpinalHDL based on Scala.

During the procedure of integrated circuit design, the decisions made at higher levels have more significant impacts on design complexity and system performance than lower levels. To tackle this issue, a pluggable and parameterized agile hardware design flow with plugin-based technique is proposed using the bottom-up strategy\cite{ref:bottom-up}.  Plugins elaborate block connection and realize corresponding function through the dependencies among services. This easy-plug hardware design method is concluded into three steps as follows: \textit{i)} Generator Definition \& Function Breakdown, \textit{ii)} Plugin Construction \& Service Instantiation, \textit{iii)} Parameter Passing \& Plugin Everything.

As illustrated in Fig.3, when a plugin is detached from top, the implicit connection between service $A''$ and service $C'$ is established. Meanwhile, the alternative function $A \to C$ adaptively replaces the previous $A \to B \to C$. \textbf{This replacement is realized via descriptions in abstraction hierarchy without any side-effects such as redundant logic and circuits.}

\subsection{DIAG Design Flow Illustration}

\begin{figure*}[tp]
	\begin{center}
		\includegraphics[width=7.1in]{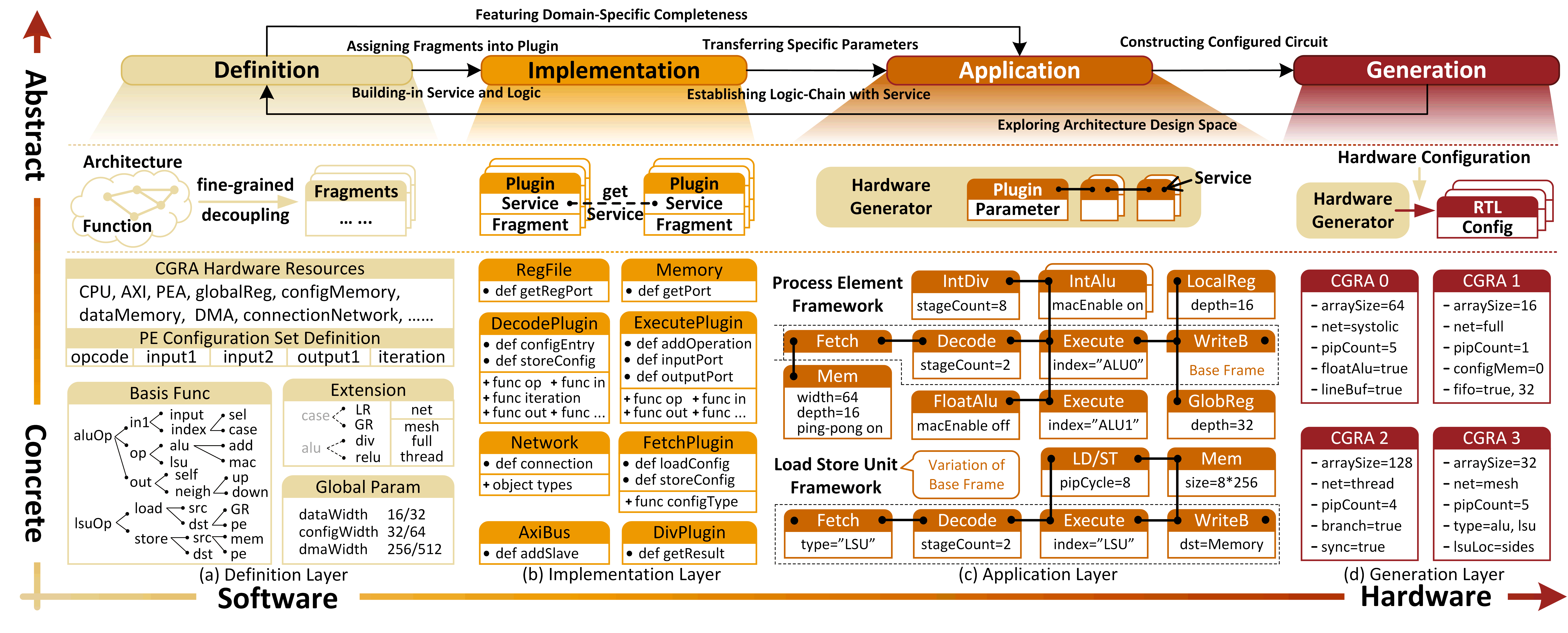}
	\end{center}
	\vspace{-0.3cm}
	\caption{DIAG paradigm. Definition, Implementation, Application and Generation layer is developed from left to right. Layer interaction, Layer abstraction and Layer instance is illustrated from top to down.}
	\vspace{-0.6cm}
	\label{fig:paradigm}
\end{figure*}

This paper refines and sublimates \textit{DIAG design flow} based on plugin technology embeded in SpinalHDL. The hardware generator can be constructed through the Definition Layer, Implementation Layer, Application Layer, and Generation Layer. The comprehensive  workflow  is depicted in Fig.\ref{fig:paradigm}.

\emph{1) Function-Oriented Definition Layer}

In the process of implementing a generator, like any hardware development, the first step is to define the specifications. What set it apart is the extraction of the commonality characteristic of processing in specific application domain. The hardware resources and architecture are seperated into three parts: \textit{i) Basic framework (i.e. basic function tree)} consists of functional fragments essential for the hardware system. \textit{ii) Extension} is a set of optional fragments for complex processing demands. \textit{iii) Parameter} is extracted from mutable hardware settings. \textbf{This specification describes a fine-grained classification of functional blocks and isolates with phsical hardware description.}

\emph{2) Plugin-Service Based Implementation Layer}

The implementation layer is where the physical description and elaboration of the definition layer introduced. \textit{Plugin} with \textit{Service} is the basic component that composes this domain. By accessing the dependencies of functions between plugins, the hardware micro-architecture is not implemented until the plugins and their parameters are establishedd, as shown in Fig.\ref{fig:paradigm} (b) and (d). \textit{All the future extensions can be structured into specific plugins and plugged in the generator.}

\emph{3) Plugin-Everything in Application Layer}

The application layer integrates plugins into a parameterized generator, as shown in Fig.\ref{fig:paradigm}(c). Tthe connections between hardware signals and their associated initialization logics inside each plugin are loaded on-demand when other plugin calls the \texttt{getService[]}. By plugging in these extensions and passing parameters, the basic framework can exhibit distinct features and functions to support complex control, precise error-checking, or even facilitate ISA modification.\textbf{In most cases, generator is designed to support a specific subset of applications within the domain to achieve considerable acceleration.}

\emph{4) Versatile Architecture from Generation Layer}

\begin{figure*}[tp]
	\begin{center}
		\includegraphics[width=7.1in]{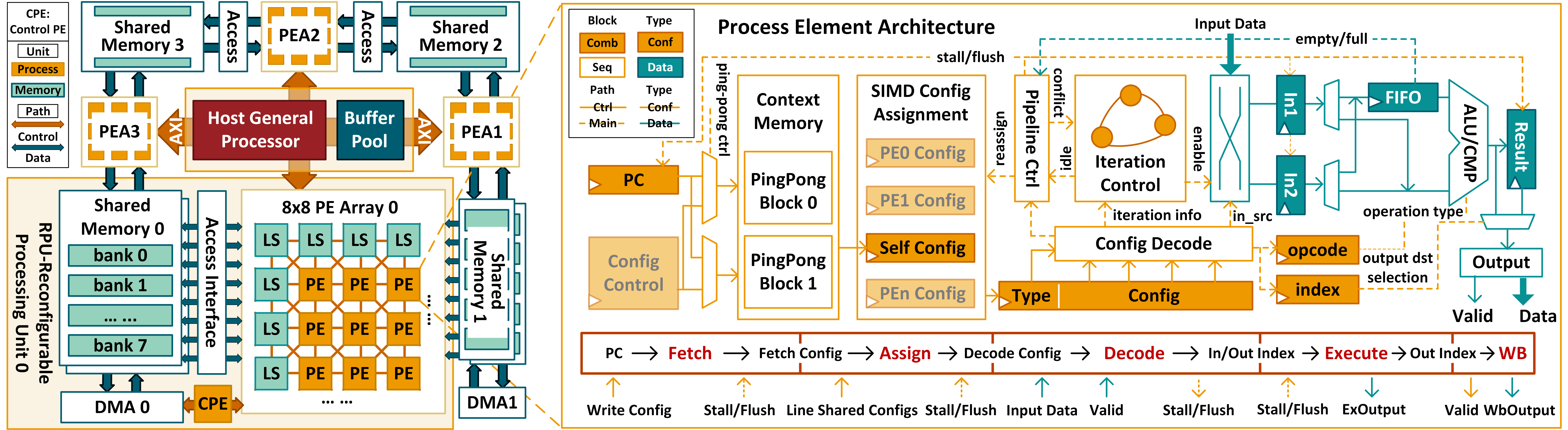}
	\end{center}
	\vspace{-0.2cm}
	\caption{Standard WindMill CGRA architecture and PE architecture inside}
	\vspace{-0.6cm}
	\label{fig:CGRA}
\end{figure*}

\textbf{Finally, the hardware circuit described in Verilog/VHDL is generated using the library defined in emergent HDLs,} such as SpinalHDL. As always, standard simulation tools and formal verification techniques is avaliable. Moreover, performance-power-area (PPA) analyzed from algorithms through emulation model can be further transformed into parameters of generator. This parameter calibration is achieved through a negative feedback loop between \textit{Generation} and \textit{Definition} Layer.

\section{WindMill CGRA Generator Implementation} \label{sec:implementation}

In general, the \textit{DIAG design flow} can be applied to arbitrary hardware development procedure using HDLs with powerful abstract syntax and high-level programming features. This paper adopts emergent HDL, SpinalHDL, to implement our cross-domain CGRA, named WindMill. The corresponding generator is called WindMill Generator. This section provides a step-by-step introduction following the four design layers.

\subsection{WindMill CGRA Architecture Definition}

\textbf{WindMill CGRA is a spatial-temporal hybrid reconfigurable architecturs supporting run-time adaptability of data-path.} In terms of spatial computing, the primary carrier of data-flow execution is process element array (PEA). This array consists of spatially parallelized PEs via an on-chip interconnection network (e.g. 2D-mesh). In terms of temporal computing, each PE within the array is programmatically configured using a program counter. However, it is important to note that each PE consists of $30\%$ control logic and $70\%$ computing logic, which is distinct with CPU. The definition of architecture is detailed below.

\emph{1) System Integration}

The WindMill architecture proposed in this paper considers the integration of host processer and reconfigurable processing unit (RPU) as the definition of CGRA, As shown in Fig.\ref{fig:CGRA}, each RPU consists of a PEA and private access memory with a parallel access interface. Four RCAs are connected on a circle, allowing partially access permission to neighbors. This arrangement executes tasks in pipeline to achieve parallelism, which not only overlaps execution time of RCAs but also provides support for larger model.

This RCA loop is integrated with Risc-V host processor named VexRiscv via AXI bus protocal. The communication procedure between host processor and accelerator has 4 steps: \textit{1)} loading configurations on PEA, \textit{2)} loading data on shared memory, \textit{3)} lauching the acceleration and \textit{4)} storing results back to host. Each stage is controlled by a register transformation table (RTT) which decodes customized instructions in CPU to PEA control signals.

\emph{2) Process Element Array}

The data concurrency feature of PEA is the key to achieving higher efficiency in data-intensive applications compared to CPU. Additionally, The configurable data-path and customized operations provide WindMill more flexibility and ways of optimization than GPU in specific domain. In the standard WindMill CGRA, general-purpose PEs (GPE) are surrounded by load store units (LSUs) which access the shared memory (SM) through parallel access interface. These special PEs (i.e. LSUs) can be configured to support both affine and non-affine access pattern. The customized PE connection network is optimized based on 2D-mesh, 1-hop, and torus topologies. Mapping, branch, loop and so on, every possible computing patterns embeded in DFG is avaliable. Shared registers provide four modes for data delivery between schedules, including line-shared, row-shared, quadrant-shared and global-shared. 

\emph{3) Process Element}

The PE serves as the fundamental unit for coarse-grained computing in the WindMill architecture. Each PE is structured into 4 pipeline stages corresponding to configuration fetch, configuration decode, execute and write back. Pipelines in PE is divided into config-flow (orange parts in Fig.\ref{fig:CGRA}) and data-flow (green parts in Fig.\ref{fig:CGRA}) to overlap schedule and execution at run-time. According to the information resolved from configuration in the Iteration Control Block, PE supports to switch control step statically and process valid operands dynamically. Additionally, WindMill CGRA provids two execution modes: single-configuration-multiple-data (SCMD) and multi-configuration-multiple data (MCMD). In SCMD mode, configurations can be shared on the same PE line, which frees up the context memory to accommodate $8\times$ configurations than MCMD.


\emph{4) Shared Memory}

The shared memory consists of two parts, banked SRAM and the parallel access interface (PAI). In a standard WindMill architecture, SRAM is seperated into 16 $256\times32$-bits banks. The round-robin arbiter is applied to PAI to arbitrate priority order of access requests from 28 LSUs. When the data model is large enough, inconsistent work between data migration and data execution causes performance bottleneck in data-flow accelerating. To address this issue, a ping-pong strategy is supported by the cooperation between the PAI and the DMA controller. The most-significant-bit (MSB) of the address is reserved after finish signal delivered by PEA periodically. Thus, the data migration from external storage and the computation in the array is overlapped.

\emph{5) Controller Process Element}

When the launch of RCA is dominated by host, the frequently layer-to-layer data movements is inefficient in accelerating multi-layered algorithms like CNN. Thus, a specific PE called CPE is introduced. The CPE is responsible for managing data and configuration migration, and controlling lauch timing. Once the host has configured the CPE, it invokes the array to calculate independently according to the defined mode. The affiliation of CPE overlaps sequential computing time on the host with concurrency computing time on the RCA. Furthermore, the design of CPE is similar with GPE except the extension of access to RTT. Implementing the CPE within the basic framework of the GPE is straightforward.

\subsection{WindMill Generator Construction in DIAG Design Flow}

The above provides a detailed definition of WindMill CGRA architecture.

\textbf{In Definition Layer}, the abstract expression of hierarchical architecture is represented in a tree structure, as shown in Fig.\ref{fig:paradigm}(a). Branches are created according to the functional fragments while leaves are initialized as \texttt{Handle[Data]} waiting for declaring required hardware types through \texttt{create early} stage in Implementation Layer. This mechanism ensures that every possible branch is avaliable but only leaves containing the required data type are finally generated.

\textbf{The Implementation Layer} involves a step-by-step refinement process, transforming abstract descriptions from the definition layer into logic plugins described in SpinalHDL. This process is completed in three stages, create config, create early and create late. Each stage follows a blocking compilation approach, where the next phase is halted until all the plugins are added in the current generator. Unlike strict coupling feature in Verilog/VHDL, every logic path is optional and the generation depends on \texttt{getService[Plugin]}.

\begin{figure}[tp]
	\vspace{-0.4cm}
	\begin{center}
		\includegraphics[width=3.4in]{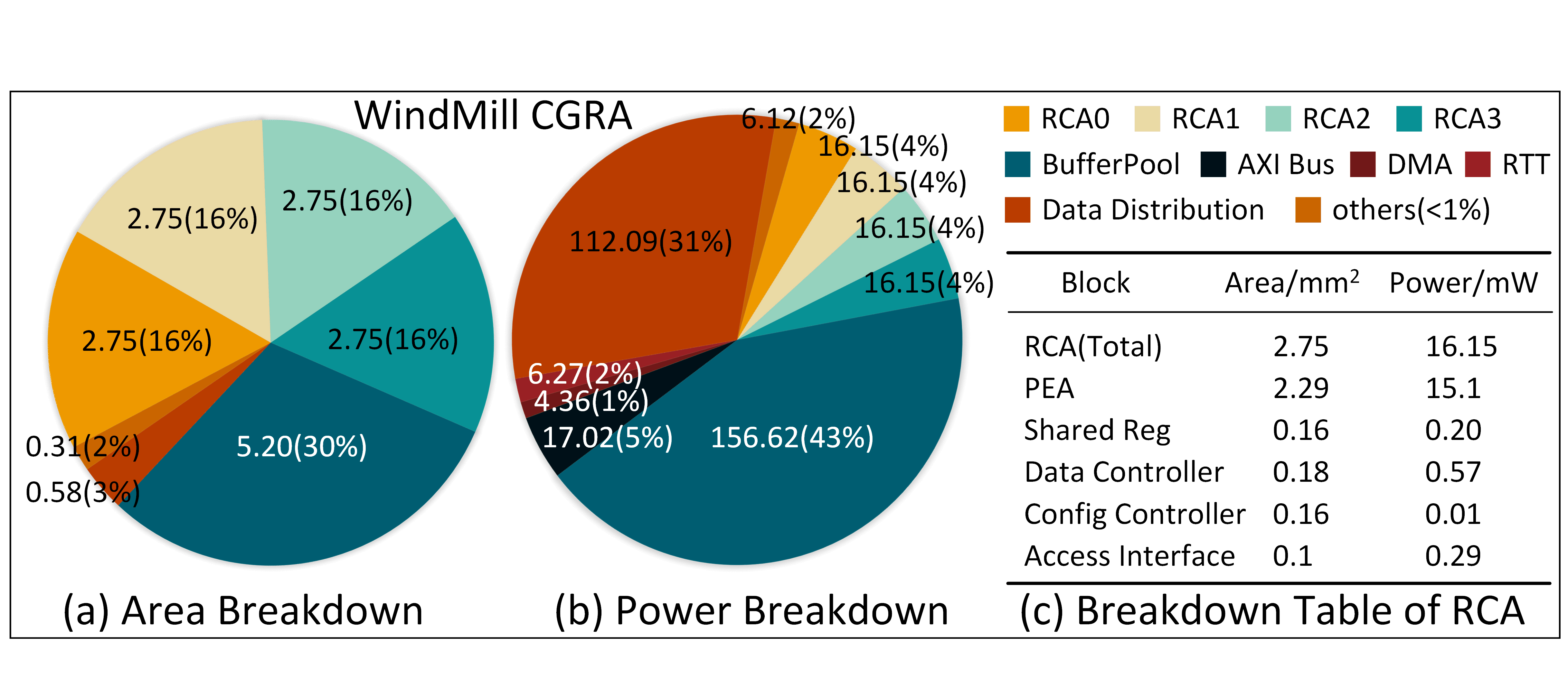}
	\end{center}
	\vspace{-0.4cm}
	\caption{Standard WindMill CGRA breakdown}
	\vspace{-0.6cm}
	\label{fig:generation}
\end{figure}

\textbf{In Application layer}, the WindMill CGRA generator is recursively constructed through the bottom-up integration of plugins. As illustrated in Fig.\ref{fig:paradigm}(c), the diversity among PE frameworks depends on the integration approach among plugins. This implementation presents significant scalability and compatibility in the SpinalHDL development platform.

\textbf{In Generation Layer}, the versatile WindMill CGRA generator is implemented based on predefined arguments and then translated into Verilog/VHDL. As shown in Fig.\ref{fig:paradigm}(d), several WindMill CGRA presets are prepared. Integrated these with stimulation procedure, not only the functionality of WindMill variant can be quickly verified, but also the computational features can be extracted, which supports further exploration for the optimal solution in specific computing domain.

\section{Experiment}

We implemented the WindMill CGRA generator using SpinalHDL, passed the pre-simulation of generated Verilog in VCS \& Verdi, and synthesized the gate-level netlist in SMIC 40nm process. Through the fully procedure from the design to the verification, advantages of WindMill CGRA and DIAG design flow is concluded into 4 aspects.

\subsection{Better Architecture Scalability}

\begin{figure}[tp]
	\begin{center}
		\includegraphics[width=3.4in]{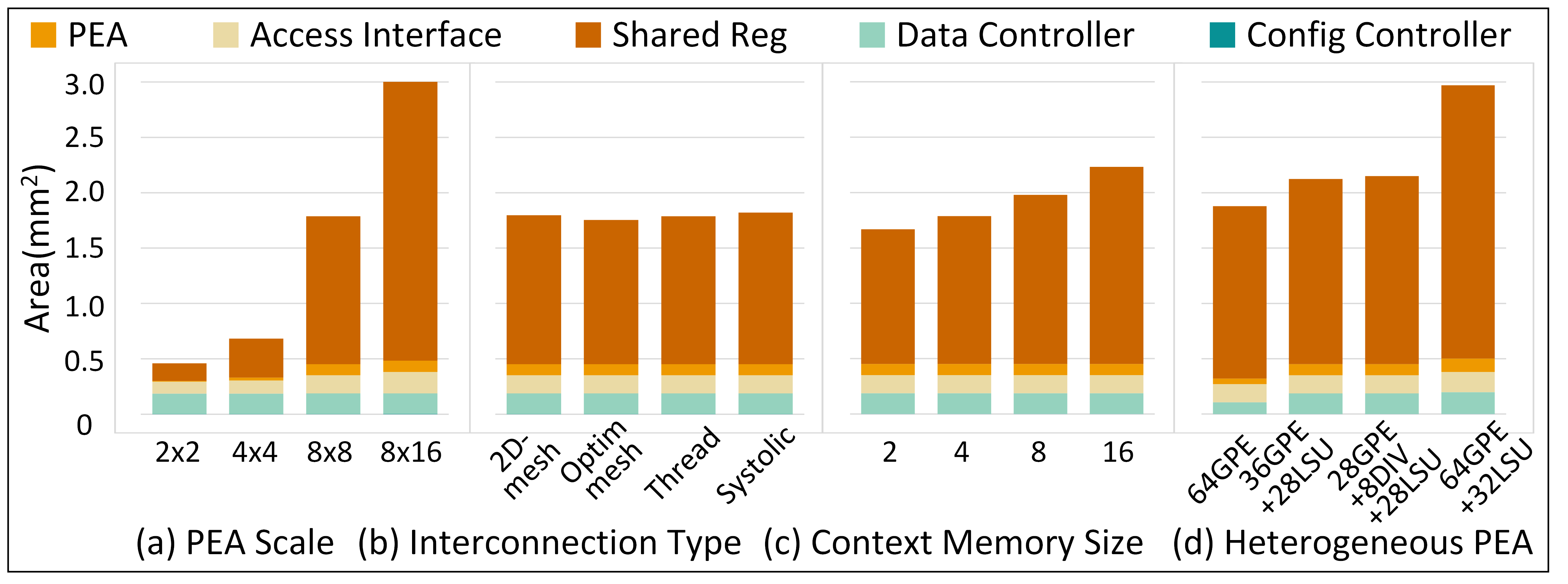}
	\end{center}
	\vspace{-0.3cm}
	\caption{Comparison of versatile CGRA}
	\vspace{-0.5cm}
	\label{fig:versatile}
\end{figure}

The variations of WindMill CGRA is generated. The parameterization capabilities of four key resources (e.g. PEA size, memory size, interconnection type) in spatial architecture are shown in the Fig.\ref{fig:versatile}. The results from Fig.\ref{fig:versatile} (a)(b) indicates that the area of generated CGRA is strongly affected by the PEA size and PE type but weakly by the interconnection topology. Furthermore, results from Fig.\ref{fig:versatile} (d) proves that the new \textit{DIAG} design flow with plugin technique presents desirable performance in easy-plug heterogeneous integration and agile productivity. \textbf{These features is beneficial to keep pace with the development of fragmental, diverse and updating domain-specific applications and further mitigate \textit{4-d Problems}.}

\section{Conclusion}

This paper proposed the WindMill CGRA generator targeted to cross-domain accelerating and implemented it in pluggable DIAG design flow. The generated hardware is able to operate at $750MHz$ and $16.15mW$ in 40nm process. Applications and algorithms from three aspects is experimented. In the case of the reinforcement learning algorithm running on the WindMill architecture, a significant performance improvement of average $200\times$ compared to CPU and $2.3\times$ compared to GPU is achieved. These results demonstrate the promising nature of the DIAG flow and the effectiveness of the WindMill architecture designed within this framework.



\end{document}